\documentclass[prl,english,showpacs,twocolumn]{revtex4-1}
\usepackage{amsfonts}

\usepackage{epsfig}
\usepackage{float}
\usepackage{latexsym}
\usepackage{graphicx}
\usepackage{amssymb}
\usepackage{amsmath}
\usepackage{color}
\begin{document}

\title{Three dimensional Lifshitz black hole and the Korteweg-de Vries
equation}
\author{E. Abdalla }
%\email{eabdalla}
\author{Jeferson de Oliveira}
%\email{jeferson}
\affiliation{Instituto de F\'isica, Universidade de S\~ao Paulo, CEP 05315-970, S\~ao
Paulo, Brazil}
\author{A. Lima-Santos}
%\email{dals@df.ufscar.br}
\affiliation{Universidade Federal de S\~{a}o Carlos, Departamento de F\'{\i}sica, CEP
13560-905, S\~{a}o Carlos, Brazil}
\author{A. B. Pavan}
%\email{alan}
\affiliation{Universidade Federal de Itajub\'a, Departamento de F\'isica e Qu\'{\i}mica, CEP 37500-903, Itajub\'a, Brazil}
\date{\today}

%%%%%%%%%%%%%%%%%%%%%%%%%%%%%%%%%%%%

\begin{abstract}
We consider a solution of three dimensional New Massive Gravity with a
negative cosmological constant and use the AdS/CTF correspondence to inquire
about the equivalent two dimensional model at the boundary. We conclude that
there should be a close relation with the Korteweg-de Vries equation.
\end{abstract}

\maketitle

%%%%%%%%%%%%%%%%%%%%%%%%%%%%%
%%%%%%%%%%%%%%%%%%%%%%%%%%%%%

%\section*{Introduction}

Recently, alternative theories of gravity have been conceived in the context
of quantum gravity, in which higher order curvature corrections appear
naturally. In such a domain, it may be even conceivable that Lorentz
invariance is not explicitly realized, though being recovered at a later
stage. This is the vein chosen by Horava \cite{horava}, to introduce a
(renormalizable) gravitational theory with a space time anisotropy, but with
a flow to general relativity at long distances. A theory of massive gravity
in three dimensions has also been considered \cite{pf,nmg}, displaying
solutions with the same kind of symmetry as the anisotropic space time
considered by Horava but in a relativistic setup.

On the other hand, the AdS/CFT relation percolated beyond the string theory
where it has been originally formulated and important developments related
AdS gravity to models of condensed matter physics \cite{hartnoll}. Special attention has been driven by the
question of superconductivity and superfluidity \cite{herzog} using this
approach.

One knows that the Korteweg-de Vries (KdV) equation describes a wide class
of physical systems \cite{scott}. Its studies have applications that spread
from hydrodynamics to condensed matter systems. The KdV equation is very
useful when one takes in account both nonlinear and dispersive effects and it
was firstly developed to study the evolution of solitary waves in shallow
water.  Some of its remarkable
characteristics are the existence of exact solutions called \emph{solitons}
and its interpretation as a completely integrable Hamiltonian system. 

In this work we establish a correspondence between a Lifshitz black hole and
a classical KdV in the spirit of AdS/CFT correspondence. From the gravity point of view, we start with a black hole solution in
three dimensions exhibiting Lifshitz scaling. It has been found in the
context of the New Massive Gravity (NMG) \cite{nmg} by Beato \textit{et al}
\cite{beato}. At the linearized level the theory is equivalent to the
unitary Pauli-Fierz theory \cite{pf} for free massive spin-2 gravitons in
three dimensions. Here we consider NMG with a negative cosmological
constant, whose action is given by
\begin{equation}  \label{action}
\mathcal{S}_{NMG}=\frac{1}{16\pi G}\int d^{3}x \sqrt{-g}\left[R-2\lambda-
\frac{1}{m^{2}}\mathcal{K}\right]\quad, 
\end{equation}
where $\lambda$ is the three-dimensional cosmological constant and $\mathcal{%
K}=R_{ab}R^{ab}-\frac{3}{8}R^{2}$. The associated Euler-Lagrange equation is
\begin{equation}  \label{euler_lagrange}
R_{ab}-\frac{1}{2}g_{ab}R=\frac{1}{2m^{2}}\mathcal{K}_{ab} - \lambda
g_{ab}\quad ,
\end{equation}
where the higher-derivative features of the theory are encoded in the tensor
$\mathcal{K}_{ab}$ (see reference \cite{nmg}). The black hole obtained by Beato {\it{et al}} is an exact solution of Eq.(\ref%
{euler_lagrange}) described by the line element
\begin{equation}  \label{metrica_lifsh}
ds^{2}=-\frac{r^{6}}{l^{6}}\ f(r)\ dt^{2}+ \frac{l^{2}}{r^{2}} \ f(r)^{-1} \
dr^{2}+r^{2}d \phi^{2}\quad ,
\end{equation}
where $f(r)=\left(1-\frac{r_{+}^{2}}{r^{2}}\right)$ and the event horizon is
denoted by $r_+$. %and the coordinates are defined to run in the intervals $%
%r\in [0,\infty[$, $t\in ]-\infty, +\infty[$ and $\phi\in[0,2\pi]$.
This solution preserves the Lifshitz scale symmetry $t\rightarrow \lambda^{3}t$, $%
\phi\rightarrow\lambda x$, $r\rightarrow\lambda^{-1}r $ with $r_{+}\rightarrow\lambda^{-1}r_{+}$.

At the spatial infinity the metric (\ref{metrica_lifsh}) reduces to
\begin{equation}  \label{metric_infinity}
ds^{2}_{r\rightarrow\infty}\rightarrow -\frac{r^{6}}{l^{6}}dt^{2}+ \frac{%
l^{2}}{r^{2}}dr^{2}+r^{2}d\phi^{2}\quad .  %\notag
\end{equation}
This asymptotic form of the metric is invariant under temporal and spacial
translation and anisotropic dilatation, $t^{\prime }=(1+3\epsilon)t$, $%
r^{\prime }=(1+\epsilon)r$ $\phi^{\prime }=(1+\epsilon)\phi$,
with 
$\epsilon$ an infinitesimal constant. This is a subgroup of the Lifshitz
Group $L(3)$ in three dimensions \cite{rongencai}
\begin{eqnarray}  \label{subgroup}
P_{\phi}=-i\partial_{\phi}, \hspace{0.3cm} H=-i\partial_{t},\hspace{0.3cm}
D=-i\left(3t\partial_{t}+\phi\partial_{\phi}+r\partial_{r}\right).  \notag
\end{eqnarray}

The propagation of a massive scalar field $\Psi(t,y,\phi)$
on the background metric (\ref{metric_infinity}) at the spatial
infinity \linebreak ($y \equiv r_{+}/r \rightarrow 0$) is given by
\begin{equation}  \label{scalar_field_infty}
\frac{d^{2}\Psi}{dy^{2}}-\frac{3}{y}\frac{d\Psi}{dy}-\frac{l^{2}\mu^{2}} {%
y^{2}}\Psi=0\quad ,  
\end{equation}
with $\mu $ 
denoting the scalar field mass. The solution near the boundary takes the
form
\begin{equation}
\Psi=\phi_{1}(t,\phi)y^{\Delta_{+}}+\phi_{2}(t,\phi)y^{\Delta_{-}},  
\end{equation}
with $\Delta_{\pm}=2\pm\sqrt{4+\mu^{2}l^{2}}$. We set $\phi_{2}(t,\phi)=0$
in order to have a finite value for $\Psi$ at the boundary, 
implying that the radial dimension of $\Psi$
is $\left[\Psi\right]=\Delta_{+}$. The result is similar to the
corresponding problem in the holographic superconductor as given in \cite%
{gubser}.

%%%%%%%%%%%%%%%%%%%%%%%%%%%%%%%%%%%%%%%%%%%%%%%%%%%
%\section*{KdV theory and its holographic description}
%\section*{KdV theory and holography}
%%%%%%%%%%%%%%%%%%%%%%%%%%%%%%%%%%%%%%%%%%%%%%%%%%%%

As in the usual search for a conformal field theory on the AdS boundary, we
looking for a field theory living at $y=0$, whose fields preserve the same
symmetries as the bulk scalar $\Psi$ and its radial dimension $\Delta_{+}$.
We claim that the best candidate is the integrable Korteweg-de Vries theory
(KdV) in (1+1)-dimensions. Letting $\Phi(t,\phi)$ be the fundamental field
at the boundary with spatial dimension $\Delta_{+}$, the KdV theory is invariant under the Lifshitz transformations and $\Phi\rightarrow\Phi^{%
\prime }=  \left(1-\Delta_{+}\epsilon\right)\Phi\quad$. We thus
get a generalization of the KdV equation
\begin{equation}  \label{gen1}
\partial_{t}\Phi + \partial^{3}_{\phi}+ \Phi^{A}\partial_{\phi}\Phi=0,
\end{equation}
where $A=\frac{2}{\Delta_{+}} $ is a constant determined imposing invariance
of the theory under the subgroup $L(3)$, in which case the bulk dimension is
$[\Phi]=\Delta_{+}$.

Using only symmetry arguments we claim that the holographic field theory
defined at the boundary $y=0$ obeys the equation of motion (\ref{gen1}).

The first two terms of (\ref{gen1}) arise naturally,
while the third one is mandatory in view of renormalization effects for the
given conformal dimension. For $\Delta_{+}=2$ we recover the usual KdV
theory. In this case the mass of scalar field $\Psi$ propagating in the bulk
is precisely $\mu^{2}_{LF}=-\frac{4}{l^{2}}$. An interesting point, is that such mass limit is smaller
than the Breitenlohner-Freedman bound \cite{bf}. One knows
\cite{hartnoll} that in three dimensions the BF bound results $%
\mu^2_{BF}=-\frac{1}{\ell^2}$, thus the Lifshitz bound is less restrictive
than BF bound, \emph{i.e.}, $\mu^2_{LF}\leq\mu^2<\mu^2_{BF}$. This new LF
bound can have relevant implications in the holographic description of
models in condensed matter.

The Klein-Gordon equation in the background geometry of Lifshitz black hole (%
\ref{metrica_lifsh}) permits considering the behaviour of the field $\Phi$
as a function of the Hawking temperature of the black hole as well as the
scaling properties of the observables in the bulk as functions of $T$. In
terms of the event horizon parameter, the temperature is given by $%
T=r_{+}^{3}/2\pi l^{4}$. We separate the Klein-Gordon equation 
with the Ansatz $\Psi(t,y,\phi)= e^{-i\omega t +i k\phi}Z(y)$ and defining $%
f(y)=1-y^{2}$. Making the redefinitions $y\rightarrow l^{1/3}\tilde{y}$, $k
\rightarrow l^{1/3}\tilde{k}$, we get
\begin{widetext}
\begin{equation}\label{kg2}
Z(\tilde{y})'' +\frac{\tilde{y}^{2}-3}{\tilde{y}f(\tilde{y})}Z(\tilde{y})' +
\left[\left(\frac{\omega}{2\pi T}\right)^{2}\frac{\tilde{y}^{4}}
{f(\tilde{y})^{2}}-\frac{l^{2}\mu^{2}}{f(\tilde{y})}-\left(\frac
{\tilde{k}}{(2\pi T)^{1/3}}\right)^{2}\right]Z(\tilde{y})=0\quad .
\end{equation}
\end{widetext}
From this equation we see that $\omega$ scales linearly with the temperature
$\omega\sim T$,while the momentum in the $\phi$ direction scales as $k\sim
T^{1/3}$, which is compatible with what we would expect from a holographic
theory with the Lifshitz scaling $t\rightarrow \lambda^{3} t$, $y\rightarrow
\lambda^{-1}y$  and $\phi\rightarrow \lambda\phi$. Given such temperature
scaling in the bulk, we ask how the fields at the boundary scale with the
black hole temperature. From the previous section, we know that $\Phi\sim
\lambda^{\Delta_{+}}$ at the boundary, where $\lambda$ is the same scaling
factor which appears in the Lifshitz transformation. In terms of the
temperature $\partial_t \sim T, \hspace{0.3cm} \partial_\phi \sim T^{1/3}$,
which implies that the fields at the boundary have the following dependence
with the black hole temperature $\Phi\sim T^{\frac{\Delta_{+}}{3}}$ .

Already briefly mentioned is the fact that in terms of the scaling
properties we see that, for an interacting theory, the last term is
compatible with all symmetries, thus, in view of renormalization effects it
becomes mandatory.

Furthermore, we address the issue of the scalar quasinormal modes of the
three dimensional Lifshitz black hole (\ref{metrica_lifsh}) in the
hydrodynamic limit. Quasinormal modes are found by solving the wave equation
for matter or gravity, under the boundary conditions of ingoing waves at
horizon and outgoing flux at the infinity for asymptotic flat spacetimes and
Dirichlet for AdS-like spacetimes. Under such conditions, the wave equation
admits solutions only for a set of discrete complex frequencies called
quasinormal frequencies. For a recent review see \cite{berticardoso}.

It is believed that in general any interacting theory can be described by
hydrodynamics in the limit of small frequencies $\omega$ and wavenumbers $k$
compared to the temperature $T$ of the system \cite{landau_fluid}. Defining
the quantities $\mathfrak{m}=\omega/2\pi T,\hspace{0.3cm} \mathfrak{q}%
=k/(2\pi T)^{1/3}$ and taking the limit $\mathfrak{q}\rightarrow 0$ \linebreak
and $\mu=0$, the equation of scalar field (\ref{kg2}) reads
\begin{equation}  \label{scalar_t}
Z^{\prime \prime }(y) + \frac{(y^{2}-3)}{y(1-y^{2})}Z^{\prime }(y) -\frac{%
\mathfrak{m}^{2}y^{4}} {(1-y^{2})^{2}}Z(y)=0\quad .
\end{equation}
In the hydrodynamic limit $\mathfrak{m}\ll 1$ we can expand the field $Z(y)$
in power series of $\mathfrak{m}$ and $\mathfrak{q}$. As we are interested
in the limit $\mathfrak{q}=0$, we can write
\begin{equation}  \label{hidro_expand}
Z(y)=(1-y^{2})^{-\frac{i\mathfrak{m}}{2}}\left[F_{0}(y)+i\mathfrak{m}%
F_{1}(y)+ \mathcal{O}(\mathfrak{m}^{2})\right]\quad .
\end{equation}
Putting back that expansion in (\ref{scalar_t}) and solving the differential
equations for $F_{0}$ and $F_{1}$ imposing causal boundary conditions, the
solution to Eq.(\ref{scalar_t}) is 
\begin{equation}  \label{solution_st}
Z(y)=A(1-y^{2})^{-\frac{i\mathfrak{m}}{2}}\left[1+\frac{i\mathfrak{m}}{2}
(1-y^{2})+\mathcal{O}(\mathfrak{m}^{2})\right], 
\end{equation}
where $A$ is a normalization constant. The Dirichlet boundary condition at
spatial infinity $Z(0)=0$, implies the relation $\omega=(4\pi T)i$,
showing that the scalar frequencies in the hydrodynamic limit are purely
imaginary. Following the AdS/CFT prescription, a black hole perturbation in
the bulk spacetime is equivalent to perturb an approximately thermal state
defined at the spacetime boundary $y=0$. In the view of linear response
theory \cite{fetter}, there is a timescale of the thermal dual system which
its perturbed state spends to return to thermal equilibrium. In particular,
such a thermalization timescale has an interpretation in terms of AdS/CFT
correspondence, namely that the characteristic damping time of the
fundamental quasinormal frequency of the black hole spacetime is related to
the thermalization timescale of the dual field theory at the boundary \cite%
{HH}. Applying such interpretation to our result, we found that the
thermalization timescale is given by $\tau= 1/4\pi T$. 
At high temperatures (compared to the wavenumber $k$ and the frequency $%
\omega$), we see that the field theory timescale is very small, indicating
that a perturbation in the two-dimensional dual thermal field theory is not
long-lived, it goes exponentially to zero. It is in accordance to the result that the 
scalar quasinormal frequency that
we found in the bulk is purely imaginary, showing that such perturbation has
no considerable oscillation stage.

Furthermore, the evidence that in some regime we have a KdV-like theory at
the boundary, as we discussed in the precedent section, is favored since one
of the most interesting features of solitonic solutions is that its
equilibrium is stable, at least in a weakly dispersing medium \cite%
{landau_kinetics}. It means that it is very difficult take the theory out
off the equilibrium. 

Another quantity that can be calculated using the AdS/CFT prescription is
the two-point correlation function associated to the massive scalar field $%
Z(y)$ in the bulk. Following the Son and Starinets recipe 
\cite{son1,son2} the momentum-space two-point function is given by
\begin{equation}  \label{op1}
\langle\mathcal{O}_{Z}(k,\omega)\mathcal{O}_{Z}(-k,-\omega)
\rangle=-2\lim_{\epsilon \rightarrow 0}F(k,\omega,\epsilon),  
\end{equation}
where $\epsilon$ is the cutoff near the boundary and $F(k,\omega,\epsilon)$
is the flux factor, which in terms of bulk metric and boundary-bulk
propagator $\tilde{G}(k,\omega,y)$ read as
\begin{equation}
F(k,\omega,\epsilon)=\lim_{y\rightarrow\epsilon}\sqrt{-g}g^{yy}\tilde{G}(-k,
-\omega,y)\partial_{y}\tilde{G}(k,\omega,y).  \notag
\end{equation}
The boundary-bulk propagator $\tilde{G}(k,\omega,y)$ is solution of
Klein-Gordon equation (\ref{kg2}) with \linebreak
$\tilde{Z}(k,\omega,y)=\tilde{G}%
(k,\omega,y)\tilde{Z}(k,\omega,\epsilon)$, where $\tilde{Z}(k,\omega,y)$ is
the massive scalar field in the Fourier space and $\tilde{Z}%
(k,\omega,\epsilon)$ is the value of the field at boundary cutoff $y=\epsilon
$. Together with Eq. (\ref{kg2}) we impose the boundary conditions: $\tilde{G%
}(k,\omega,\epsilon)=1$ and $\tilde{G}(k,\omega,1)$ being finite.
Explicitly, the solution is
\begin{equation}  \label{green1}
\tilde{G}(k,\omega,y)=\left(\frac{y}{\epsilon}\right)^{ \Delta_{+}}\left[%
\frac{y^{2}-1}{\epsilon^{2}-1}\right]^{\frac{b}{2}}\frac{H \left[0,\alpha,b,-%
\frac{b^{2}}{4},c,y^{2}\right]}{H\left[0,\alpha,b,- \frac{b^{2}}{4}%
,c,\epsilon^{2}\right]},  \notag
\end{equation}
where $H$ stands for the confluent Heun function and \linebreak $\alpha=\sqrt{%
4+\mu^{2}l^{2}}$, $b=il^{4}\omega/r_{+}^{3}$, $c=%
\alpha^{2}/4+l^{2}k^{2}/4r_{+}^{2}$. Taking the expansion of $\tilde{G}%
(k,\omega,y)$ and $\tilde{G}(-k,-\omega,y)$ near $y=\epsilon$ and putting
back in Eq.(\ref{op1}) we get

\begin{widetext}
\begin{equation}\nonumber
\langle\mathcal{O}_{Z}(k,\omega)\mathcal{O}_{Z}(-k,-\omega)
\rangle=\frac{2r_{+}^{4}}{l^{4}}\lim_{\epsilon \rightarrow 0}\left\{
 \frac{y^{2\Delta_{+}}\epsilon^{-2\Delta_{+}}\bar{H}\left[\cdots,y^{2}\right]}
{\bar{H}\left[\cdots,\epsilon^{2}\right]H\left[\cdots,\epsilon^{2}\right]}
\left[\frac{\Delta_{+}}{y}(y^{2}-1)+\frac{b}{y^{2}}\right]\left[H
\left[\cdots,y^{2}\right]+\partial_{y}H\left[\cdots,y^{2}\right]\right]\right\},
\end{equation}
\end{widetext}
where $\bar{H}$ refers to the confluent Heun function with $b\rightarrow -b$.
Expanding the Heun's function up to sixth order, it is straightforward to
see that this results only in contact terms in the expression for the
two-point correlation function, without any term which could give rise to
correlations between points with both spatial and temporal separation. Such
localized terms in the correlator expression confirm our claim that the
boundary theory is solitonic.

Consider again the scalar field in the Lifshitz black hole $\Psi
(t,y,\varphi )=q^{1-\frac{\alpha }{2}}(q-1)^{\frac{b}{2}}X(q)\mathrm{e}%
^{-i\omega t+ik\varphi }$. We get the confluent Heun equation%

\begin{equation}
X^{^{\prime \prime }}+\left( \frac{1-\alpha }{q}+\frac{1+b}{q-1}\right)
X^{^{\prime }}-\frac{b^{2}}{4(q-1)}X=0  \label{che}
\end{equation}%
where $\alpha $, $b$ are given above and $q=y^{2}$. From (\ref{che}) it is
possible to get a Hamiltonian structure \cite{Slavy} in order to get the
confluent Painlev\'{e} $P^{VI}$ equation \cite{Ince}
\begin{eqnarray}
q_{tt} &+&\frac{1}{2}\left[ \frac{1}{q}+\frac{1}{q-1}\right] q_{t}^{2}+\frac{%
1}{t}\left[ \frac{(1-\alpha )(q-1)}{q}+1\right] q_{t}  \notag \\
&+&\frac{1}{2t^{2}}\ Q_1 -\frac{1}{t}\ Q_2 -\frac{1}{2}\ q(q-1)(2q-1) =0,  \label{pain}
\end{eqnarray}%
as the equation of motion with

\begin{eqnarray}
Q_1&=&\left[ \frac{q(1+b)^{2}}{(q-1)}-\frac{(q-1)(1-\alpha )^{2}}{q}\right]\ \textrm{and} \nonumber\\
Q_2&=&q(q-1)\left[ 1+\frac{b(b-2)}{2}-2+\alpha \right] \nonumber.
\end{eqnarray}
The {\small ARS} conjecture \cite{ARS} states that reductions of solitonic
equations is a Painlev\'{e} equation. In particular, the reduction of the
KdV equation is the Painlev\'{e} $P^{II}$ \cite{Fokas}. Equation (\ref%
{pain}) contains the five remaining Painlev\'{e} equations \cite{Ince}. Thus,
we can derive from it the Painlev\'{e} $P^{II}$ in order to put the KdV
equation in the bulk of the Lifshitz black hole studies.  Additionally this
realization permits us speculate wether a possible interpretation of the quantum
theory living on the boundary could be an Ising model in 2 dimensions. Indeed,
the results obtained by McCoy \emph{et.al} \cite{mccoy} and Ablowitz
\cite{Fokas} establish a connection between KdV equation and Ising model
through the Painleve equation. Therefore, a general relationship between Lifshitz
black hole, KdV equation and Ising model could, in principle, be realized.

(\textit{Concluding remarks}). In this work we have shown through different
arguments and results a close relation between a 3-dimensional Lifshitz
black hole and a classical KdV theory. Another result that corroborates
this interpretation is presented by Troncoso \cite{troncoso}.

We believe that the mass bound has implications in all holographic models in condensed matter.
In principle, the
holographic superconductors can condensate more easily in
asymptotically Lifshitz models than asymptotically AdS models.

A question not answered up to now in the context of the AdS/CFT relation is
the role played by integrable two dimensional theories. As a matter of
fact, many can be seen as a perturbation of conformally invariant models. At
the classical level, some are indeed conformally invariant but integrable at
the quantum level \cite{abdbergweizs}, displaying also quantum
conformally invariant solutions, as is the case of the chiral Gross-Neveu
models in two dimensions \cite{abdallasant}. The KdV equation is, to our
knowledge, the first case where the relation can be established. A bulk
gravitational theory in asymptotically Lifshitz space equivalent to further
integrable models might shed new light in both the structure of
integrability as well as in the Gauge/Gravity duality itself.

(\textit{{Acknowledgments}}). This work has been supported by FAPESP, FAPEMIG and CNPq, Brazil.

%%%%%%%%%%%%%%%%%%%%%%%%%%%%%%%

\end{document}